\documentclass[letterpaper]{article}

\usepackage[T1]{fontenc}

\usepackage{geometry}
\usepackage{setspace}
\usepackage{amsthm}
\def\ps@pprintTitle{%
 \let\@oddhead\@empty
 \let\@evenhead\@empty
 \def\@oddfoot{}%
 \let\@evenfoot\@oddfoot}
\makeatother
\makeatletter
\newcommand*{\rom}[1]{\expandafter\@slowromancap\romannumeral #1@}

\makeatother

\usepackage{braket}
\usepackage{dcolumn}
\usepackage{xcolor}
\usepackage[utf8]{inputenc}
\usepackage{amsmath,scalefnt}
\numberwithin{equation}{section}
\usepackage{amssymb}
\usepackage{dsfont}
\usepackage{physics}
\usepackage{color}
\usepackage{graphicx}
\usepackage{mathtools}
\usepackage{overpic}
\usepackage[toc,page]{appendix}
\usepackage{chemfig}
\usepackage{ccycle}
\usepackage{comment}
\usepackage{textcomp}
\usepackage{gensymb}
\usepackage{graphics}
\usepackage{epsfig}
\usepackage[T1]{fontenc}
\usepackage{soul}
\usepackage[symbol]{footmisc}
\usepackage{mathrsfs}
\usepackage{hyperref}
\usepackage{float}
\usepackage{subcaption} 
\usepackage[style = chem-acs]{biblatex}
\usepackage{graphicx}
\usepackage{float}
\newfloat{scheme}{htbp}{los}
\floatname{scheme}{Scheme}
\floatname{chart}{Chart}
\newfloat{graph}{htbp}{loh}

\usepackage{chemformula} 
\usepackage[version = 4]{mhchem} 

\usepackage{authblk}
\author[1]{Eeshan Ketkar}
\author[2]{Swarnabha Chattaraj}
\author[1,2]{Koichi Tanaka}
\author[1]{Samy Kouidri}
\author[1,2]{Supratik Guha*}
\affil[1]{Pritzker School of Molecular Engineering, University of Chicago, Chicago, Illinois, 60637, USA}
\affil[2]{Materials Science Division, Argonne National Laboratory, Lemont, Illinois 60439, USA}

\title{Direct-Write Chemical Vapor Deposition}
\date{*Email: guha@uchicago.edu}

\begin{document}

\maketitle

\begin{abstract}
  Direct-write pattering, using chemical vapor deposition through a nozzle, enables localized film growth using high-purity precursors standard to the electronics industry. Yet, achieving sub 100 $\mu\mathrm{m}$ feature control has been limited by nozzle design and unquantified precursor surface diffusion. Here, we report a direct-write chemical vapor deposition platform that employs a 10 $\mu\mathrm{m}$ tip diameter glass micro-nozzle mounted on a scanning stage above a heated substrate in ultra-high vacuum. Using this approach, we demonstrate the continuous single-pass writing of 500 $\mathrm{nm}$ thick metallic aluminum lines with a 30 $\mu\mathrm{m}$ linewidth on titanium nitride-coated silicon. This capability enables maskless, localized growth under systematically varied deposition conditions on a single substrate. To analyze the resulting deposit profiles, we introduce a transport framework coupling Knudsen gas emission from the micro-nozzle with surface diffusion to extract the effective precursor diffusion length, $L_d$. $L_d$ captures the combined impact of precursor surface mobility, residence time, substrate temperature ($165^{\circ}\mathrm{C}$ to $205^{\circ}\mathrm{C}$), growth duration, and nozzle height, contracting from 17.4 $\mu\mathrm{m}$ down to 3.2 $\mu\mathrm{m}$ at elevated temperatures. These results establish $L_d$ as a quantitative parameter for evaluating laterally confined chemical vapor growth and predicting feature resolution limits. This approach provides a practical methodology to measure and quantify precursor surface transport in localized film synthesis.
\end{abstract}

\section*{Keywords}

Chemical Vapor deposition, Interfacial and surface science, nanofabrication, Additive manufacturing, Area Selective Deposition


\section{Introduction}
"Direct-Write" deposition of materials has received increased attention over the past decade\cite{https://doi.org/10.1002/smsc.202300121,doi2024nanofabrication,doddapaneni2025beyond,porrati2023site,chrisey2000power,puers2001fabrication,spanu2021development,matsui2000three,winkler2018high,fowlkes2016simulation,winkler20193d,skoric2019layer,barth2020precursors,plakhotnyuk2024atomic,https://doi.org/10.1002/smtd.202101546,de2020gas}, due to their potential for creating complex three-dimensional (3D) monolithic structures, curved surfaces, and custom repairs and tailoring of circuits. These methods enable the laterally confined delivery of material to a substrate either through a nozzle or slit that is brought close to the substrate surface, or via a reactive deposition area enabled by a focused energetic beam. Methods have included physical\cite{doi2024nanofabrication} and atomic layer deposition\cite{plakhotnyuk2024atomic,https://doi.org/10.1002/smtd.202101546} (PVD and ALD); deposition of inks\cite{chrisey2000power,doddapaneni2025beyond,kim2006direct}, nanoparticles\cite{doi2024nanofabrication,doi:10.1126/science.aan2958}, and electron/ion triggered reactive deposition\cite{porrati2023site,puers2001fabrication,matsui2000three,winkler2018high,winkler20193d,fowlkes2016simulation,skoric2019layer,barth2020precursors,jungwirth2022focused,shorubalko2016direct,dubner1991role}. All of these target spatially localized mask-free material growth, enabling additive fabrication with micron-scale (or larger) precision. A description of their pros and cons may be found in a previous work\cite{doi2024nanofabrication}. Among these, chemical vapor methods such as chemical vapor deposition (CVD) and ALD offer compatibility with a wide range of standard, ultrapure precursors from the semiconductor industry and broad material deposition choices\cite{pierson1999handbook,stringfellow1999organometallic,foroughi2011atomic,barry2021chemistry}. 
\par
Direct write ALD methods have been introduced by Plakhotnyuk\cite{plakhotnyuk2024atomic} and Kundrata et. al.\cite{https://doi.org/10.1002/smtd.202101546}, using nozzles structured as a central circular delivery line with diameter ranging from 20 to 80 $\mu\mathrm{m}$ surrounded by concentric rings for vacuum and secondary precursor delivery (diameter ranging from 1000 $\mu\mathrm{m}$ to 1160 $\mu\mathrm{m}$). These nozzles are capable of depositing features with lateral dimension of the order of a few 100 $\mu\mathrm{m}$ and deposition rates  of ~0.35 $\mathrm{nm}$ per cycle for $\mathrm{TiO_{2}}$ deposition. An early-stage prototype of a direct-write CVD tool operating at atmospheric pressure (air or an argon environment) was demonstrated by Spanu et. al.\cite{spanu2021development}. Tungsten trioxide ($\mathrm{WO_{3}}$) was deposited using a nozzle held at a substrate distance $\geq$ 50 $\mu\mathrm{m}$, followed by annealing the deposit at $500^{\circ}\mathrm{C}$. This method demonstrated $\mathrm{WO_{3}}$ features with a lateral resolution of about 1 $\mathrm{mm}$. Such methods, while being influenced by the nozzle diameter and substrate-to-nozzle distance, can also be significantly affected by material redistribution due to lateral diffusion. There have been different theoretical studies of thin film crystal growth \cite{stringfellow1999organometallic,Madhukar01011988,madhukar_growth_1993,larsson2022chemical,venables1983nucleation,Ohring2002,zhang1997atomistic,grillo2020area,doi:10.1021/acs.chemmater.5c02530,almeida2015calculation,drahushuk2019fundamental}, and a clear assessment of many of the underlying atomistic mechanisms influencing adsorption-desorption and surface migration may be found in the monograph by Madhukar et al.\cite{Madhukar01011988}. However, all of these studies target large area growth where the incident material flux on the substrate is uniform across the substrate surface. Studies of crystal growth processes for laterally confined deposition (as enabled by nozzles), have been limited\cite{doi2024nanofabrication,doddapaneni2025beyond}.
\par
In this work, we report on the design and operation of an ultra-high vacuum (UHV)-compatible direct-write chemical vapor deposition (DW-CVD) platform that utilizes 10 $\mu\mathrm{m}$ tip diameter glass micro-nozzles mounted on a 3-axis piezoelectric stage with 10 $\mathrm{nm}$ out-of-plane positioning precision and active tilt compensation. Using this system, we demonstrate continuous, single-pass writing of 500 $\mathrm{nm}$ thick metallic aluminum lines with $30~\mu\mathrm{m}$ linewidths on titanium nitride-coated silicon. To quantify the nanoscale prescursor transport phenomena governing feature broadening during spatially confined growth, we establish a theoretical framework that couples Knudsen gas-phase emission from the micro-nozzle with continuum surface diffusion and line-of-sight shadowing. By comparing experimentally measured 3D deposit profiles against model predictions across systematic variations in substrate temperature ($165^\circ\mathrm{C}$ to $205^\circ\mathrm{C}$), deposition time, and nozzle-to-substrate separation, we extract the effective precursor surface diffusion length, $L_d$. This approach reveals how precursor residence times, mobility, and incorporation kinetics evolve under localized flux delivery, establishing $L_d$ as a quantitative benchmark parameter for predicting physical feature resolution limits across direct-write CVD,PVD and area selective growth processes.

\section{Results and discussion}

\subsection{Direct-Write chemical vapor deposition (DW-CVD) system}
\begin{figure*}[t]
    \includegraphics[width=1\textwidth]{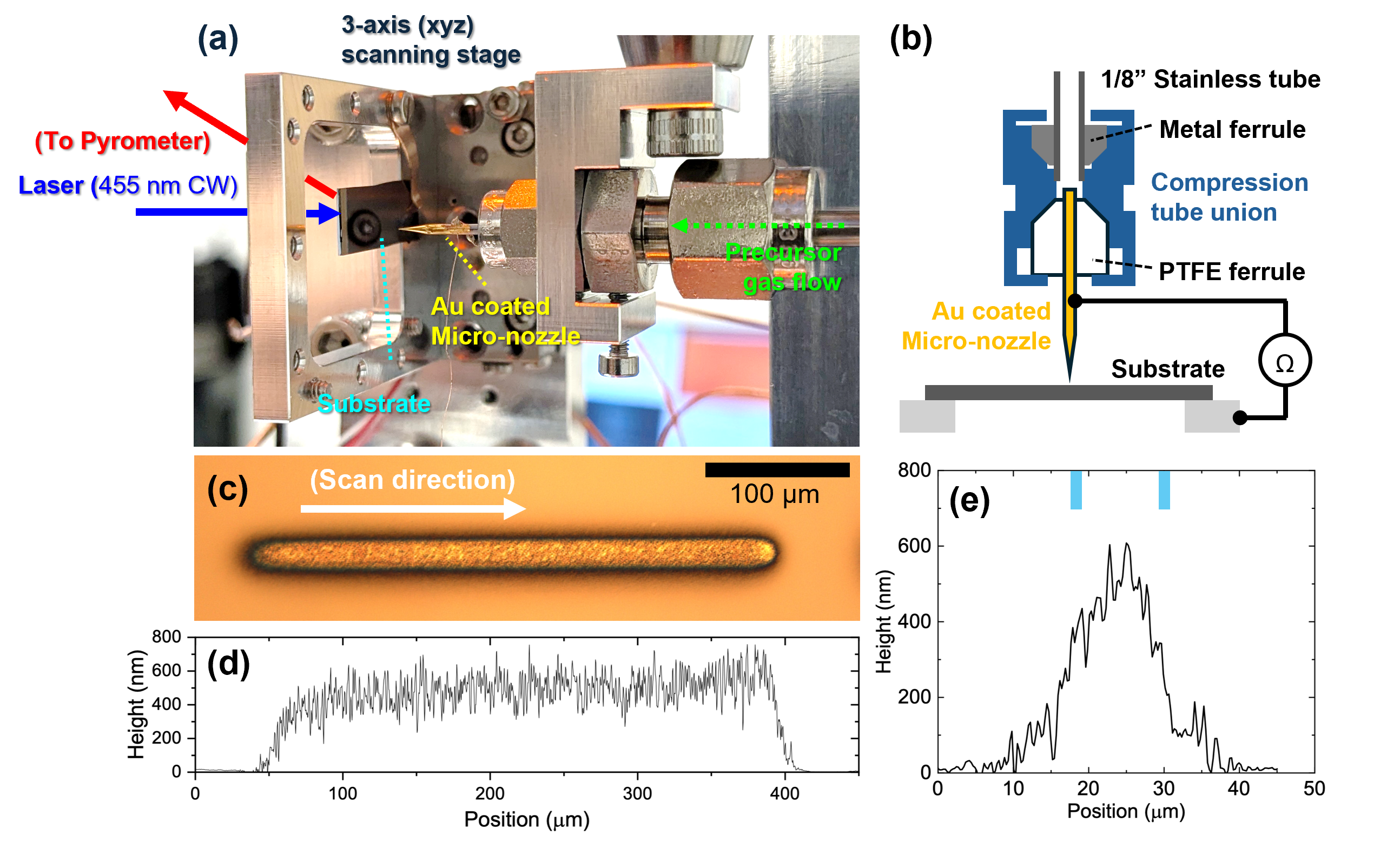}
    \caption{
Overview of the Direct-Write chemical vapor deposition system. (a) Schematic illustration of the Direct-Write CVD experimental setup, (b) Nozzle configuration for the "touchdown" surface detection method, (c - f) representative results of localized, scanning DW-CVD of aluminum (Al) deposited by scanning DW-CVD method. (c) an optical micrograph of an Al line, height profile of the Al line along (d) longitudinal and (e) cross-sectional direction. Blue indicators in (e) represent the relative position of the nozzle tip with respect to the scale of the height profile.}
    \label{fig:schematics}
\end{figure*}

In this subsection, we introduce the design of our Direct-Write chemical vapor deposition (DW-CVD) system. The  system is built in a UHV chamber pumped by a rotary pump and a turbo molecular pump (~230 l/s). Typical base pressure is $\sim5\times10^{-8}$ Torr.  A photograph of the deposition chamber and a schematic diagram of the DW-CVD system are shown in Fig\,\ref{fig:schematics}(a) and (b).  As shown in Fig.\,\ref{fig:schematics}\,(a), the deposition components consist of: (i) a glass micro-nozzle that has a tapered tip and (ii) a 3-axis scanning stage with a substrate holder. Precursor vapor is delivered to the substrate surface via the tapered glass micro nozzle (internal diameter ($\mathrm{d}$) of 10 $\mu\mathrm{m}$ at the outlet and $\mathrm{d} = 0.66$ $\mathrm{mm}$ away from the tapered end were used for the studies reported here; purchased from World Precision Instrument Inc.). The inlet end is connected to a 1/8 inch stainless steel gas tubing via a compression-type tube fitting (Swagelok Inc.) with a Teflon (PTFE) ferrule. This tubing leads to a standard canister based CVD precursor delivery system. The precursor flow and resulting chamber pressure is controlled via a leak valve. Prior to loading into the system, the exterior of the glass nozzle is coated with gold (Au) by sputter-deposition for detecting nozzle-surface contact electrically (this is further described in the following paragraph). Note that the nozzle tip is electrically isolated from the stage and sample and connected to a digital multimeter (DMM, 6500 from Keithley Inc.) by a Copper (Cu) wire and Indium (In) solder. (See Fig.\,\ref{fig:schematics}\,(b))
\begin{figure*}[t]
    \includegraphics[width=0.9\textwidth]{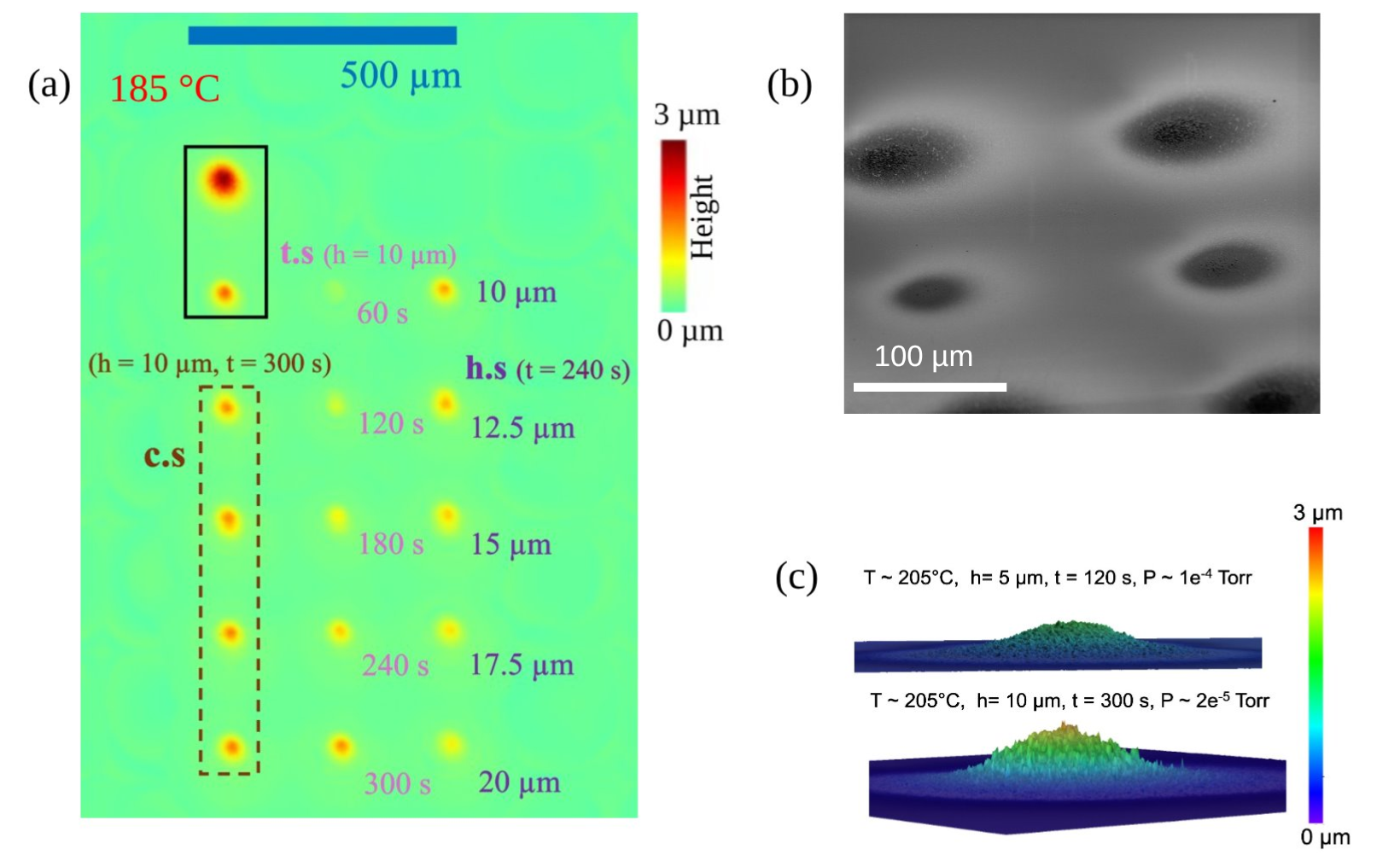}
    \caption{(a)Representative contour plot (height in color bar) of our deposited matrix (at $185^{\circ}C$) used for $L_\mathrm{d}$ analysis. Dummy depositions have been highlighted in the black boxes. c.s (in brown and dashed outline), t.s (pink) and h.s (purple) represent the consistency scan, time scan and height scan columns, respectively. The deposition conditions associated with each element of a column have been labeled with the colors associated with that column type. (b) Tilted SEM images of a few spot depositions of the array depicted in (a). (c) 3D profiles of two spots deposited at two different chamber pressures, nozzle to substrate heights and deposition times from two different deposition experiments.}
    \label{fig:array}
\end{figure*}
\par
A piece of single-side polished Si(100) wafer (approximate dimensions are $4 \times 15 \times 0.3$ mm\textsuperscript{3}), ultrasonically cleaned in acetone followed by 2-propanol for 5 minutes each, is mounted on an aluminum alloy sample holder. The holder  has a rectangular through-hole at its center to enable heating of the sample by a laser. The sample holder is loaded onto a 3-axis piezoelectric inertia drive stage assembly, which moves the substrate holder in plane (x and y) and out-of-plane (z) directions. The resolution of each axis is 0.75 $\mu\mathrm{m}$ and 10 $\mathrm{nm}$ along the in-plane and out-of-plane directions, respectively. The substrate is heated by a continuous wave (CW) laser ($\lambda$ = 455 $\mathrm{nm}$) from the exterior of the chamber through a fused-silica viewport. The substrate temperature is monitored using a pyrometer aligned to the laser incident spot. 
\par
Prior to deposition, surface detection and tilt compensation sequences are carried out for precise control of the substrate-nozzle distance across the substrate surface. In this procedure, the substrate is moved toward the nozzle until the resistance between the substrate and the nozzle (monitored by the digital multimeter) drops below 100 M$\Omega$, indicating nozzle-substrate contact initiation ("touchdown"). This enables determining the accurate position of the substrate surface to within an error of $10~\mathrm{nm}$. We repeat the touchdown procedure at three different locations that are 500 $\mu$m apart from one another, and in this fashion are able to calculate the tilt of the substrate along the in-plane drive axes (i.e. x and y). Note that typical surface slope values ($\Delta z/\Delta x, \Delta z/\Delta y$) obtained experimentally in this fashion are approximately 0.1-0.5 \%. Prior to each deposition sequence, we carry out the single touchdown process to detect the substrate surface and set it as the origin of the out-of-plane axis, i.e. z = 0. Following the touchdown, the sample stage is lowered along the z axis while the translation distance is recorded by the piezo inertia stage. During the scanning deposition, the distance between the nozzle and the original substrate surface is maintained at a fixed value and the offset in out-of-plane direction (z) caused by the substrate tilt is compensated to keep a constant substrate-nozzle distance across the scanning range. It is important to note that the nozzle-substrate distance reported is that between the flat substrate surface and nozzle prior to deposition; note that the local working distance decreases as the deposition front advances towards the nozzle tip (we do not compensate for the change in local working distance).
\par
All depositions in this experiment were carried out on either bare Si (100) or $\sim$22-nm-thick \ce{TiN} layer deposited on Si (100) .  For Al deposition using DW-CVD we used a metal-organic precursor dimethylethylaminealane (DMEAA, C\textsubscript{2}H\textsubscript{5}N(CH\textsubscript{3})\textsubscript{2}·AlH\textsubscript{3}). 0.5 M solution of DMEAA in toluene (purchased from Millipore-Sigma Inc.). The canister filled with the DMEAA/toluene solution is  connected to the gas delivery line that leads the precursor vapor to the substrate surface through the glass nozzle. The canister is heated by a ribbon heater at $\sim80^{\circ}\mathrm{C}$. The flow of the precursor vapor is manually controlled by a leak valve equipped in the gas delivery line to achieve a constant chamber pressure $p = 2\sim6\times10^{-5}$ Torr. Representative characteristics of an Al line prepared using the DW-CVD system is shown in Fig.\,\ref{fig:schematics}\,(c)-(e). The Al line was prepared by scanning DW-CVD on bare Si(100) substrate with a chamber pressure $p = 6\times10^{-5}$ Torr at a substrate temperature $T_{s} = 189^{\circ}\mathrm{C}$  using a glass nozzle with a tip diameter $d_{tip} = 10 \mu\mathrm{m}$ placed at a nozzle tip-substrate distance $h = 0.7\mu\mathrm{m}$. During the scanning deposition, the substrate was moved \textit{in plane} such that relative position of the nozzle moved along the scanning direction indicated in (c) at a constant rate of $2\mu\mathrm{m}/\mathrm{s}$ ($1\mu\mathrm{m}/\mathrm{step}, 0.5 \mathrm{s}/\mathrm{step}$). Using the developed scanning capability, a line-shaped feature of Al is successfully deposited by DW-CVD for the length of 250 $\mu\mathrm{m}$ as shown in Fig.\,\ref{fig:schematics}\,(c). It is important to note, that our setup allows us to print a 500\,$\mathrm{nm}$ thick line in a single pass, unlike previous work\cite{plakhotnyuk2024atomic,https://doi.org/10.1002/smtd.202101546}, which requires multiple passes along a scanning direction to print a line shape with sub ~30\,$\mathrm{nm}$ thickness using ALD. The height profile along the longitudinal direction shown in Fig.\,\ref{fig:schematics}\,(d) indicates that the top surface of the Al line has an areal root mean square surface roughness $S_{\mathrm{q}}$ = 0.177 $\mu\mathrm{m}$, caused by a polycrystalline grain morphology (Fig.\,\ref{fig:schematics}\,(e)). The mean height level of the deposit is constant along the scanning direction. From the cross-sectional height profile (Fig.\,\ref{fig:schematics}\,(f)), we measure the FWHM of the line to be approximately 15 $\mu\mathrm{m}$, which is 50\% larger than $\mathrm{d_{tip}}$. This observation indicates that the lateral transport of materials (precursor molecules or their products) in the gas phase and/or on the surface significantly affect the resolution of the DW-CVD method and needs to be considered.

\par
To estimate $L_{\mathrm{d}}$, we compare a series of experimentally deposited profiles with simulations based on the theoretical model described later in the paper. We deposit a series of Al "spots" in  $p = 2\times10^{-5}$ Torr at three substrate temperatures on our PEALD \ce{TiN} coated Si (100) substrate in the following sequence: $205^{\circ}\mathrm{C}$, $185^{\circ}\mathrm{C}$ and $165^{\circ}\mathrm{C}$. Each "spot" is a consequence of CVD deposition of Al precursor through a nozzle ($\mathrm{d_{tip}} = 10~\mu\mathrm{m}$) positioned at a certain nozzle-to-substrate distance and for a set duration. This leads to an axially symmetric deposition profile or "spot". For each deposition temperature we create a matrix of spots, each spot corresponding to a specific deposition condition. An image of our deposition matrix is shown in Fig. \ref{fig:array}.  There are three columns of spots in each deposition matrix. The first column is the consistency scan sequence (denoted as ``cs''). It consists of four nominally identical spot depositions performed with nozzle height $h=10~\mu\mathrm{m}$ and deposition time $t=300~\mathrm{s}$. This column is used to test the fidelity of the CVD process by measuring the reproducibility of the profile shape and extracted $L_{\mathrm{d}}$. The second column is the time scan sequence (``ts'') and is used to study the variation of $L_{\mathrm{d}}$ as a function of deposition time. Here, the nozzle height is held fixed at $h=10~\mu\mathrm{m}$ for deposition times of $t=300,240,180,120,60~\mathrm{s}$. The third column is the height scan sequence (``hs''), used to study how $L_{\mathrm{d}}$ varies with nozzle height from the substrate surface. The nozzle heights used are $h=10,12.5,15,17.5,20~\mathrm{\mu}m$ and the deposition time is held fixed at $t=240~\mathrm{s}$. In addition, each matrix has two test depositions at the top left corner carried out at $h=25~\mu\mathrm{m}$, $t=600~\mathrm{s}$  and $h=10~\mu\mathrm{m}$, $t=300~\mathrm{s}$. These are used to test the stability of the temperature and chamber pressure prior to the subsequent spot depositions. All matrices are deposited at identical precursor delivery line pressures of $2$ Torr and chamber pressures of approximately $2\times10^{-5}$ Torr. Substrate detection and tilt calibration is carried out before the deposition of each matrix. Each spot within a matrix is separated by $200~\mu\mathrm{m}$, while each deposition matrix is separated by approximately $500~\mu\mathrm{m}$ on the substrate. Following deposition of the matrix of Al spots, the 3D deposit profiles are measured using laser confocal microscopy (Evident Scientific LEXT OLS5000). Further details on these measurements as well as SEM images of the deposit microstructure can be found in the Supporting Information section S1 and S2.

\begin{figure*}[htbp]
    \centering

    \begin{subfigure}[t]{0.31\textwidth}
        \centering
        \caption{}
        \includegraphics[width=\linewidth]{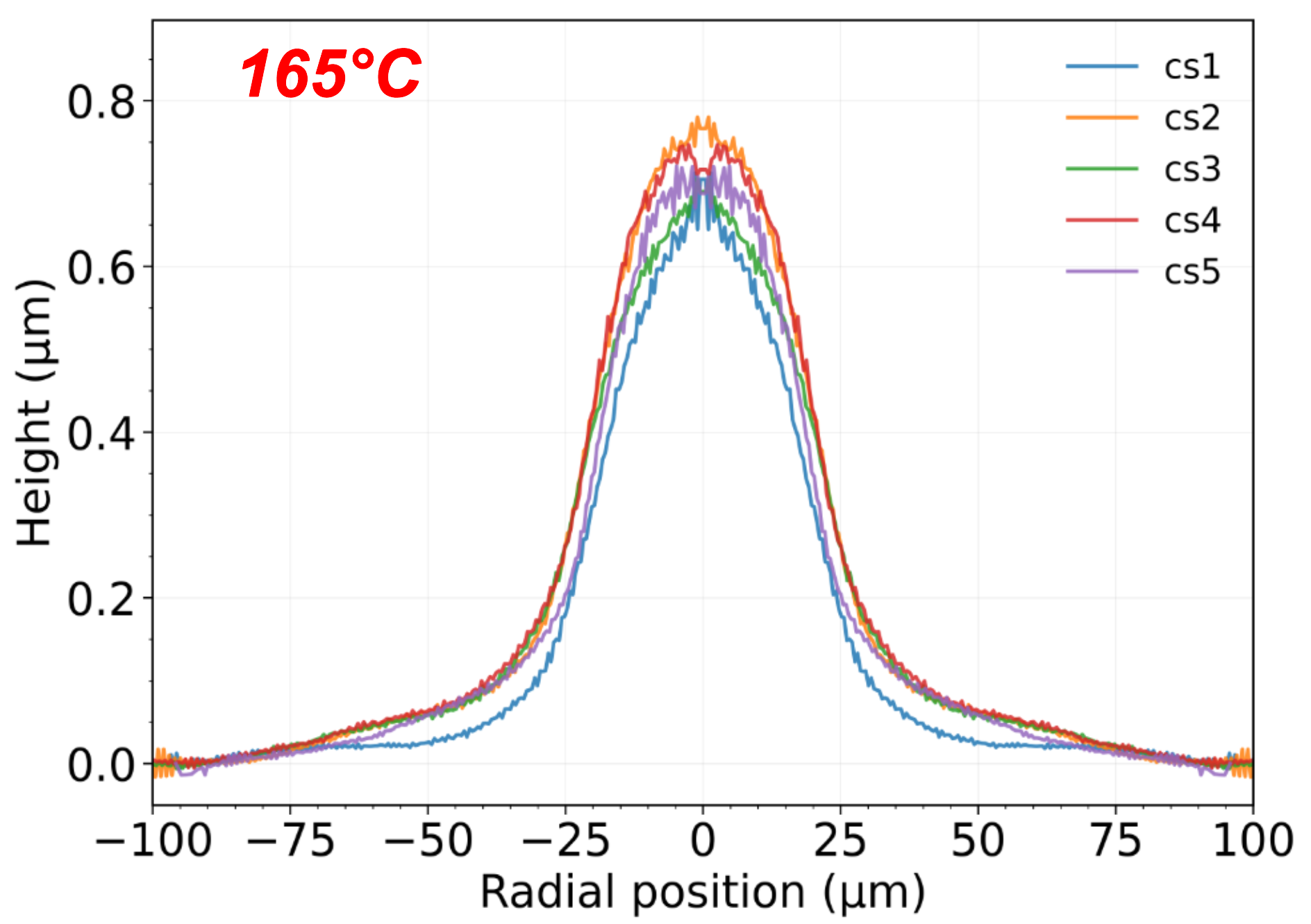}
        \label{fig:grid_a}
    \end{subfigure}
    \hfill
    \begin{subfigure}[t]{0.31\textwidth}
        \centering
        \caption{}
        \includegraphics[width=\linewidth]{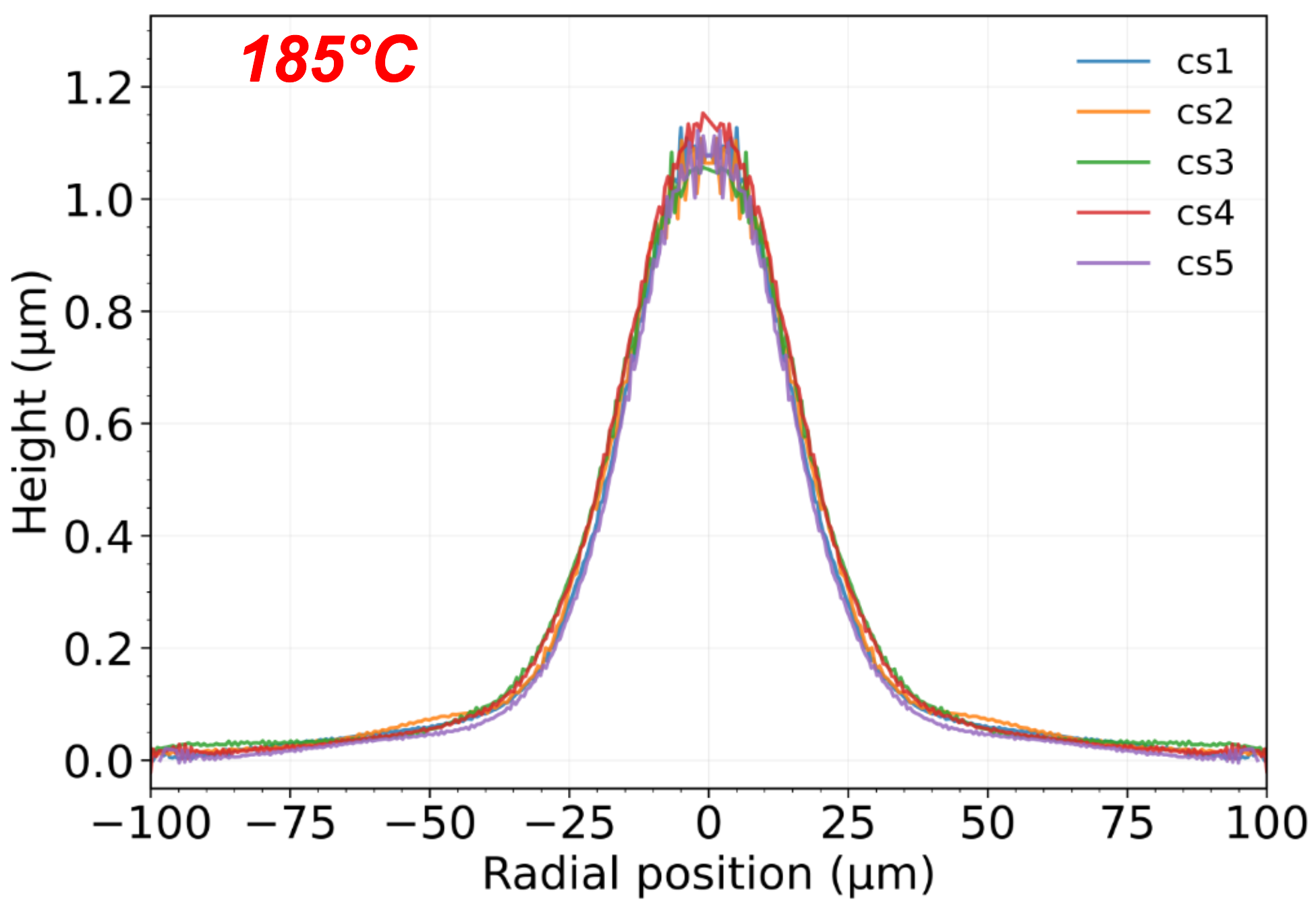}
        \label{fig:grid_b}
    \end{subfigure}
    \hfill
    \begin{subfigure}[t]{0.31\textwidth}
        \centering
        \caption{}
        \includegraphics[width=\linewidth]{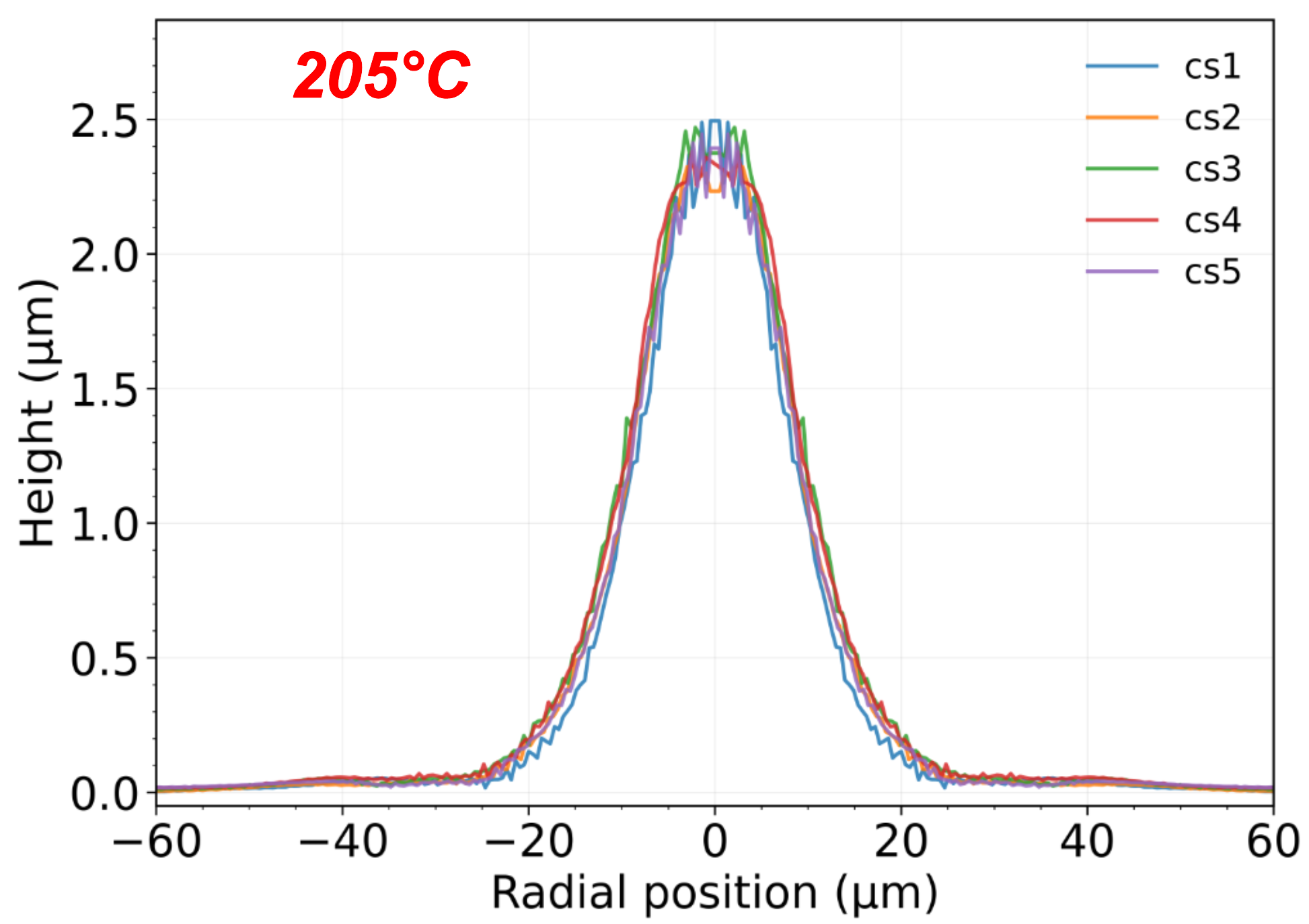}
        \label{fig:grid_c}
    \end{subfigure}

    \vspace{0.1cm}

    \begin{subfigure}[t]{0.31\textwidth}
        \centering
        \caption{}
        \includegraphics[width=\linewidth]{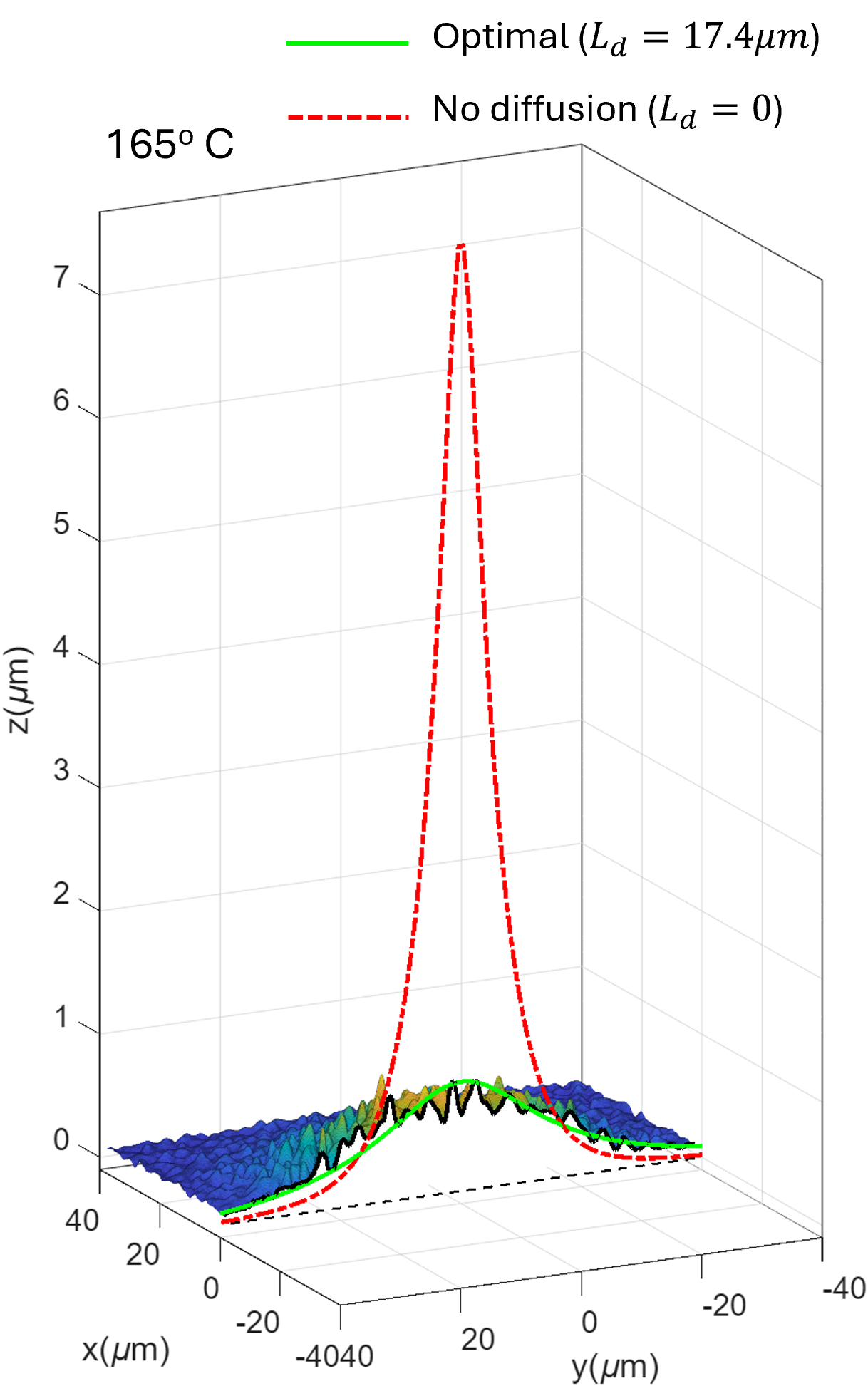}
        \label{fig:grid_d}
    \end{subfigure}
    \hfill
    \begin{subfigure}[t]{0.31\textwidth}
        \centering
        \caption{}
        \includegraphics[width=\linewidth]{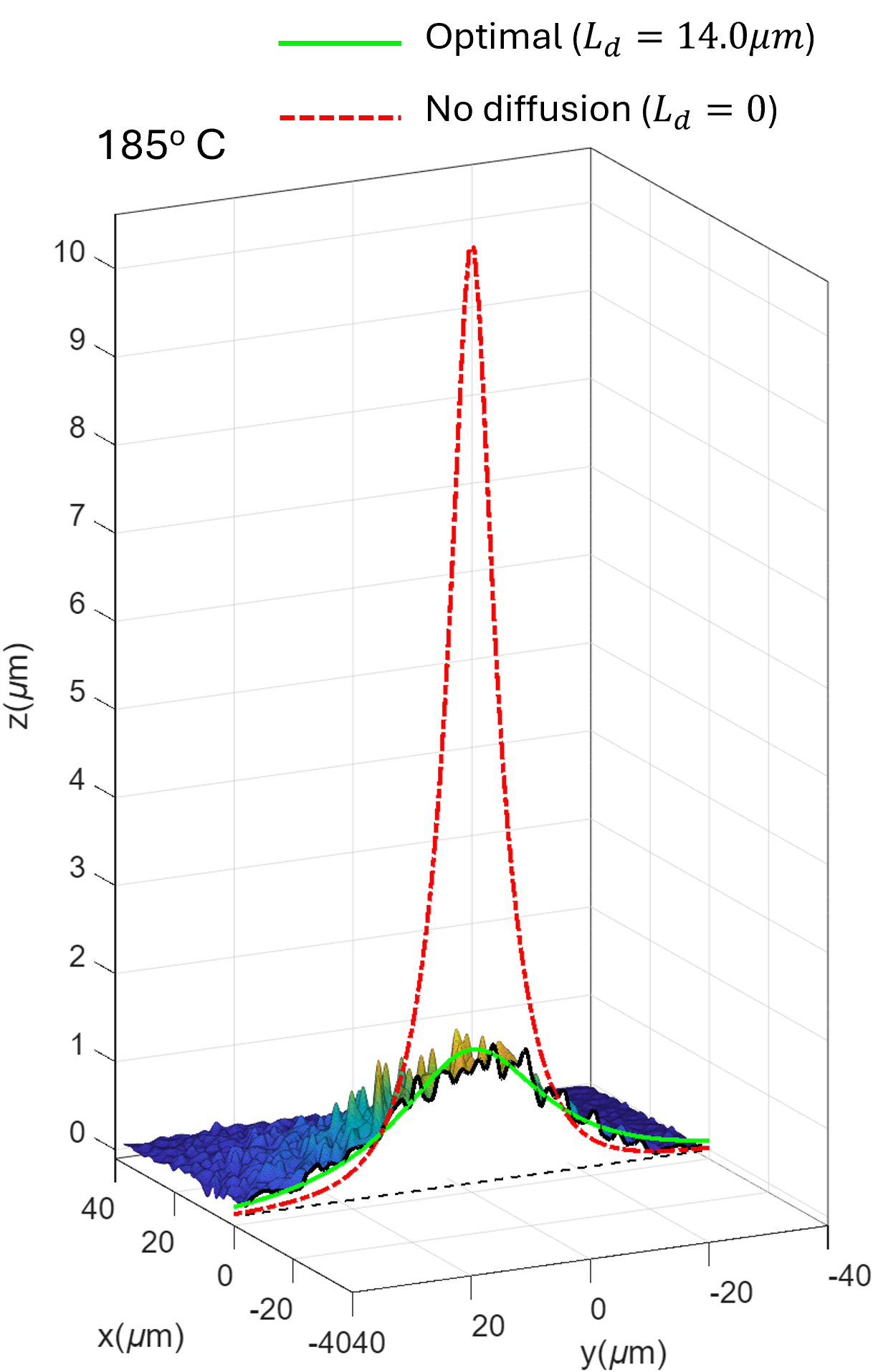}
        \label{fig:grid_e}
    \end{subfigure}
    \hfill
    \begin{subfigure}[t]{0.31\textwidth}
        \centering
        \caption{}
        \includegraphics[width=\linewidth]{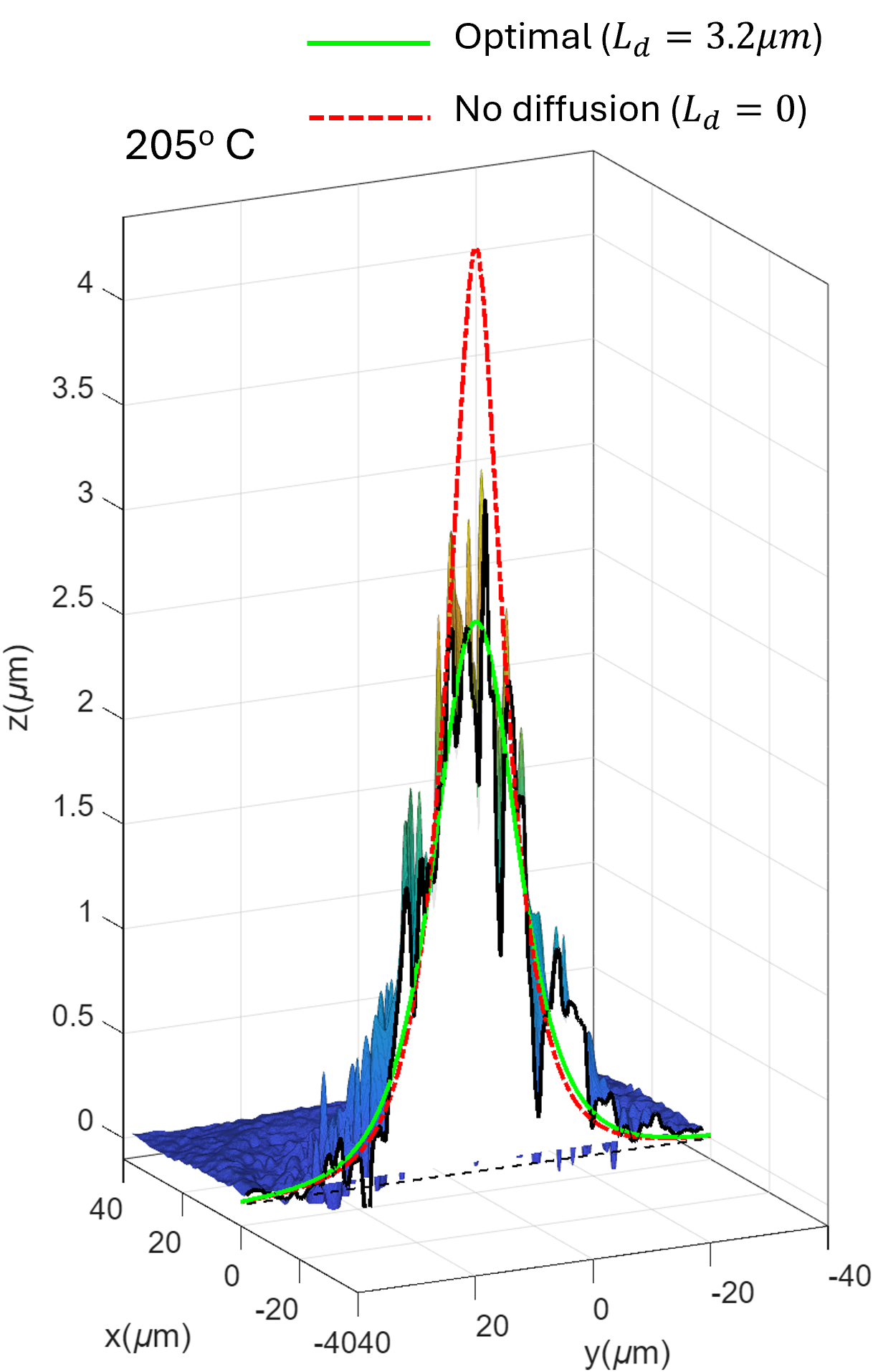}
        \label{fig:grid_f}
    \end{subfigure}

    \vspace{0.1cm}

    \caption{%
    Panels (a)-(c) show the radially averaged deposit profiles (plotted in radial coordinates) from the consistency scans at 
    $165^{\circ}\mathrm{C}$, $185^{\circ}\mathrm{C}$, and $205^{\circ}\mathrm{C}$, respectively, demonstrating run-to-run reproducibility. 
    For representative cases at each temperature, panels (d)-(f) show the 3D experimental data corresponding to the consistency scan (plotted in cartesian coordinates) with the XZ section shown. In overlay we show in green line the best fit with optimal $L_{\mathrm{d}}$ determined using the approach outlined in Section \ref{eld}. To highlight the effect of $L_{\mathrm{d}}$ on the model profile, we show as dashed red curve the model profile with no diffusion ($L_{\mathrm{d}} = 0$). In computing the $L_{\mathrm{d}} = 0$ curve, we have kept the same evolution time as the optimal green curve. This ensures that the total deposit volume is same between the green and the red curves. Evidently the deposit volume is more concentrated at the center of the profile in absence of diffusion, and more spread out in the tail region with diffusion present. 
    }
    \label{fig:cs}
\end{figure*}

\subsection{Spot deposition analysis}
We first establish run-to-run reproducibility. In Fig.\,\ref{fig:cs}\,(a), (b) and (c) we  show the experimentally measured radially averaged profiles of the deposit corresponding to substrate temperatures of $165^{\circ}\mathrm{C}$, $185^{\circ}\mathrm{C}$ and $205^{\circ}\mathrm{C}$, respectively for the consistency scan column. The different colors represent 5 different spot depositions carried out at each of the aforementioned substrate temperatures, implying reproducible outcomes. We also see that the width of the deposit shows a consistent increase with decrease in temperature. This behavior has been investigated further by quantitatively extracting the optimal parameter $L_{\mathrm{d}}$ with the continuum growth model 
including (1) Estimation  of the incident flux of adatoms ejected from the nozzle onto the substrate, and (2) Inclusion of the surface migration of the adatoms and growth. We include the first by assuming a Knudsen flow regime. The surface migration effects are included using a continuum surface diffusion model similar to previous work\cite{fujii1995theory}, but generalized for a curved deposit surface and taking into account shadowing effects. Comparing the predictions of the model with the experimentally obtained deposition profiles enables the extraction of $L_{\mathrm{d}}$ from the measured profile. Details of the model and estimation of $L_{\mathrm{d}}$ are discussed in \ref{model_ref}.  
\\
We show in Fig.\,\ref{fig:cs}\,(d), (e) and (f) the 3D fit of our model to the experimental data corresponding to the forth consistency scan (cs4) as a representative, deposited at $h = 10\,\mu\mathrm{m}\,,t = 300\,\mathrm{s}\,,p=2\times10^{-5}\,\mathrm{Torr}$. 

Here the 3D experimentally obtained profile is shown as the colored surface with the YZ section exposed. The best fit from the model with the optimal diffusion length parameter is shown as the green curve. As a reference we also show as the dashed red curve the case with no diffusion ($L_{\mathrm{d}}=0$). 
\\
In computing the $L_d=0$ curve, we keep the same deposition time as the green optimal $L_d$ curve, resulting in a narrower, and taller deposit profile compared to the experimentally measured profile (Fig.\,\ref{fig:cs}\,(d), (e) and (f) indicating that the migration of mobile adsorbed species constitutes an important part of the growth kinetics and can determine the ultimate resolution of the method.  Practically, this allows us to now estimate the optimal $L_{\mathrm{d}}$ for various experimental control parameters which has been lacking in similar studies on this topic\cite{doi:10.1021/acs.chemmater.5c02530,grillo2020area,almeida2015calculation,mandal2024selectivity}. 

\begin{figure}
    \centering
    \includegraphics[width=0.9\linewidth]{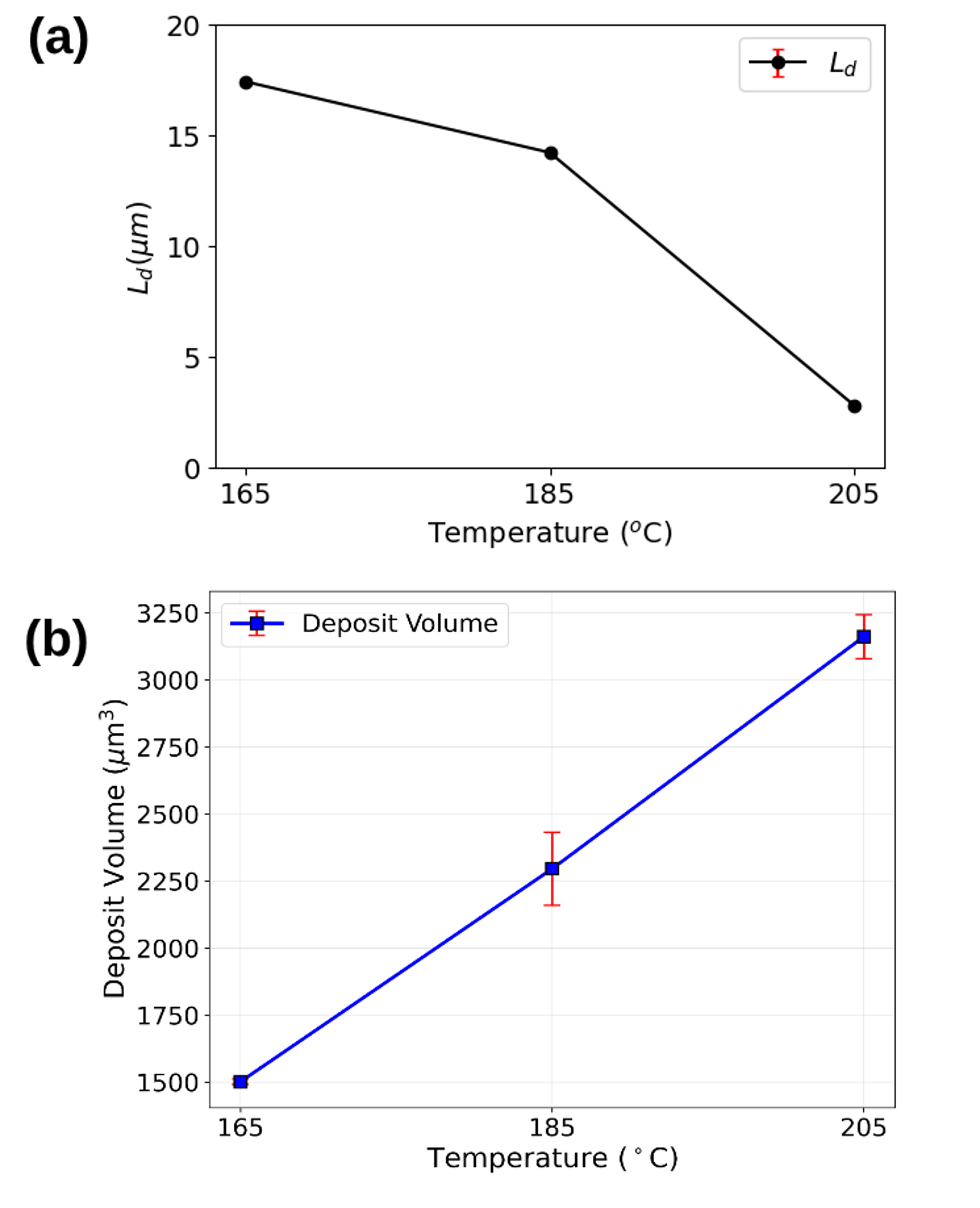}
    \caption{(a) Optimal $L_{\mathrm{d}}$ as a function of temperature plotted from consistency-scan fit data. Red dots represent the optimum $L_{\mathrm{d}}$ fit. Green bars represent the confidence intervals. (b) Mean deposition volume of all consistency-scan deposits as a function of substrate temperature.}
    \label{fig:ss_temperature}
\end{figure}
\begin{figure*}
    \centering
    \includegraphics[width=1.01\linewidth]{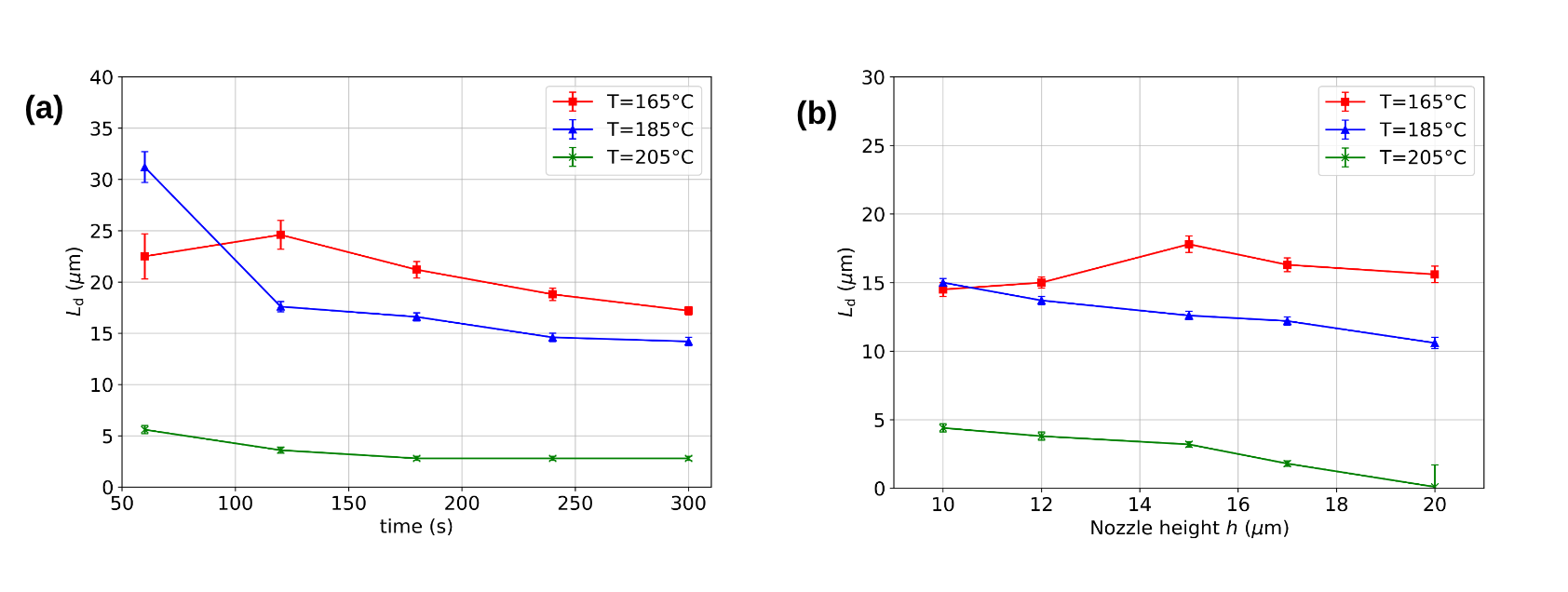}
    \caption{(a) Optimal $L_{\mathrm{d}}$ as a function of deposition time $t$ from the time-scan dataset, $h = 10\,\mu m$. (b) Optimal $L_{\mathrm{d}}$ as a function of nozzle height from the height scan-scan dataset, $t = 240\,s$.}
    \label{fig:vs_time}
\end{figure*}

\subsubsection{Effect of substrate temperature}\label{vsT}
As shown in Fig.\ref{fig:ss_temperature}(a), the extracted $L_{\mathrm{d}}$ is found to decrease with increasing temperature.  The diffusivity   $D_s$ is expected to increase with temperature, so from Eqn. \ref{num_dem} we conclude that  $\tau_s=1/k_{\mathrm{tot}}$ must be decreasing with increasing temperature.  As has been noted earlier $k_{tot}$ represents the combined effects of the precursor dissociation/incorporation rate and the precursor desorption rate ($k_{tot} = k_{des} + k_{incorp}$). To determine whether dissociation/incorporation or desorption dominates this trend, we plot the mean deposit volume as a function of substrate temperature from the consistency scan column data at $h=10~\mu\mathrm{m}$, $t=300~\mathrm{s}$ at deposition pressure of $2 \times 10^{-5}$ Torr (Fig.\,\ref{fig:ss_temperature}(b)). Since the deposition volume increases with increasing substrate temperature, the results reveal dominance of precursor dissociation and incorporation into the deposit over precursor loss via desorption. Because of increased reaction rates at higher temperatures the precursor molecules have a higher probability per unit time to react and incorporate into the growth surface, leading to a shorter residence times and therefore a smaller $L_{\mathrm{d}}$.

\subsubsection{Effect of deposition time}\label{vst}
 Fig.\,\ref{fig:vs_time}(a) shows an example of extracted $L_{\mathrm{d}}$ vs time trends for the deposition matrix ($h = 10\,\mu\mathrm{m}$) at different temperatures. As expected, the decrease in $L_{\mathrm{d}}$ with increased substrate temperature is observed for all times. Additionally there is a 25-50 percent decrease in   $L_{\mathrm{d}}$ with increased deposition time. This trend may be related to a nucleation effect. During CVD growth the precursor molecules diffuse outward and away from the zone of impingement (defined by the nozzle geometry). The substrate surface "seen" by the precursor molecules (before they are reactively incorporated or desorbed) is either the bare Si/TiN surface that has not yet been overgrown with Al, or an Al coated surface, or a combination of the two. This effectively implies two different surface diffusivities depending upon the surface coverage encountered by the precursor molecule. As the deposition proceeds and the spatial extent of Al coverage on the surface expands due to the diffusion tails, the precursor molecule's interaction will overwhelmingly be with an Al covered surface. These Al sites on the substrate surface act like catalytic traps which will inhibit the movement of the Alane precursor, while causing it to decompose and incorporate into the cluster as an Al atom\cite{kodas2008chemistry}. This would cause the estimated $L_{\mathrm{d}}$ to decrease as time progresses. 

\subsubsection{Effect of nozzle height}\label{vsH}
The height-scan analysis at $t=240\,\mathrm{s}$ is shown in Fig.\,\ref{fig:vs_time}(b). Consistent with Fig.\,\ref{fig:ss_temperature}(a), $L_{\mathrm{d}}$ reduces with increasing substrates temperature at all substrate to nozzle distances. At all three substrate temperatures, the extracted $L_{\mathrm{d}}$ decreases slightly as the nozzle-substrate separation $h$ is increased. The physical mechanism responsible for this behavior is unclear. We speculate this may be related to increased thermal coupling between the substrate and nozzle.

\subsubsection{Surface diffusion and impact on resolution}
 These studies, carried out  under varied temperature, time and height conditions, indicate that the effective surface diffusion length of the precursors can vary significantly from a few microns to a few tens of microns. The precursor gas delivery profile, which depends upon nozzle diameter and nozzle to substrate distance, will determine resolution for larger features, however it appears that for the case of Al CVD, the surface diffusion length will limit minimum lateral resolution to a few microns. While these measurements have been made for Al, we have presented a general experimental method and theoretical model that will allow diffusion length (and therefore estimate of lateral resolution) extraction for the CVD deposition of other materials.  

\subsection{Conclusions}
In this work, we have demonstrated a direct-write chemical vapor deposition (DW-CVD) platform capable of maskless, localized film synthesis using industry-standard high-purity precursors. By integrating 10 $\mu\mathrm{m}$ glass micro-nozzles with 10 $\mathrm{nm}$ out-of-plane piezoelectric positioning and electrical touchdown surface sensing, the platform enables precise, continuous single-pass writing of 500 $\mathrm{nm}$ thick metallic aluminum features with 30 $\mu\mathrm{m}$ linewidths. To evaluate and control surface-mediated feature broadening, we developed a kinetic transport model coupling Knudsen gas emission with continuum surface diffusion and shadowing effects. Comparing experimental 3D deposit profiles against pure ballistic flux simulations ($L_d = 0$) confirms that precursor surface migration, rather than geometric flux delivery alone, limits lateral feature resolution. Systematic parameter sweeps reveal that the effective diffusion length $L_d$ contracts from $17.4~\mu\mathrm{m}$ at $165^\circ\mathrm{C}$ down to $3.2~\mu\mathrm{m}$ at $205^\circ\mathrm{C}$, driven by thermally accelerated precursor dissociation and incorporation kinetics over desorption loss. Furthermore, the time-dependent reduction of $L_d$ highlights the role of catalytic site generation as metallic aluminum nucleates and spreads on the substrate. This combined experimental and theoretical framework establishes $L_d$ as a fundamental, morphology-derived metric that bridges microscale gas delivery with local surface kinetics, providing a predictive framework to optimize resolution limits across direct-write, area-selective, and localized vapor-phase growth processes.  
\\

\section{Methods : Theoretical model of DW-CVD process}\label{model_ref}

\begin{figure*}[t]
    \includegraphics[width=0.4\textwidth]{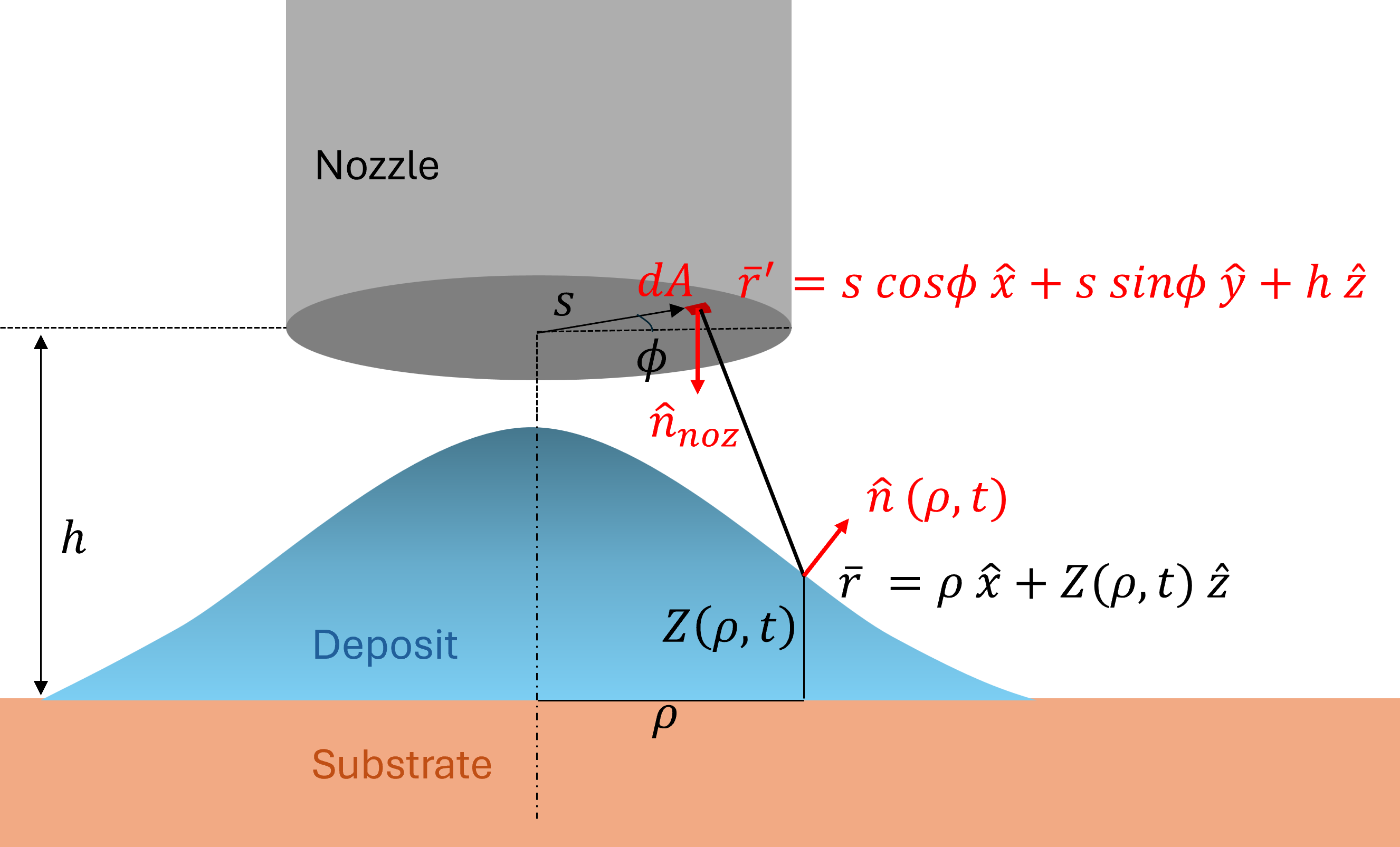}
    \caption{
    Axisymmetric model to estimate the growth profile. Here $s$,$\phi$ are coordinates denoting the location of an emitting area element $dA$ on the surface of the nozzle-aperture. The radially symmetric deposit profile is represented by the surface $Z(\rho,t)$.}
    \label{fig:model}
    
\end{figure*}

\subsection{Estimating flux from the nozzle to the substrate}
We assume that the nozzle  behaves like a simple effusion aperture whose angular emission profile is given by the Knudsen equation, valid for the case of free molecular flow. 
The Knudsen approach assumes that $\lambda \gg \mathrm{h}$, where $\lambda$ is the mean free path of the gas molecules and $\mathrm{h}$ is a characteristic distance, taken to be the nozzle to substrate distance. $\lambda$ was calculated using the collision diameter of toluene estimated in prior work\cite{sharipov2014theoretical,alvarez2013cartography}, see Supporting Information S3. Since we know the gas line conductance and the measured inlet pressure, we estimate the precursor gas pressure to be ~1 $\mathrm{Torr}$ near the nozzle mouth. This is an overestimate since the true pressure is likely lower due to the vacuum chamber pumps evacuating the nozzle area since chamber pressure is $ 2 \times 10^{-5}\,\mathrm{Torr}$. At 1 $\mathrm{Torr}$ pressure we estimate $\lambda \sim 27\,\mu\mathrm{m}$, i.e. $\lambda > h$. Furthermore, MOLFLOW simulations confirm qualitative agreement between the nozzle flux profile and a Knudsen cosine distribution~\cite{kersevan2009introduction,kersevan2019recent}. This is further described in Supporting Information section S3. We assume an axially symmetric deposition, consistent with our circular nozzle aperture and the measured nozzle tilt $<1^\circ$ in our setup described above. 
In the Knudsen case, the angular flux distribution is proportional to the cosine of the angle with respect to the surface normal of the source. As indicated in Fig. \ref{fig:model}, we represent the deposit profile at a time $t$ as $Z(\rho,t)$, where $\rho$ is the in-plane radial distance from the center. Owing to the cylindrical symmetry, it is sufficient to only consider the slice where $\rho$ is along the $X$ axis. In that case, the surface normal vector can be denoted as:

\begin{equation}
    \label{nhat}
    \hat{n}(\rho,t) = \frac{-\partial_{\rho} Z \, \hat{x} + \hat{z}}{\sqrt{1 + (\partial_{\rho} Z)^2}}
\end{equation}

Now, we consider an area element $\Delta A$ on the nozzle surface at a position $\mathbf{r}'= s \cos\phi \hat{x} + s \sin \phi \hat{y} + h \hat{z}$, and a surface normal $\hat{\mathbf{n}}_{nozzle} = -\hat{z}$ based on the configuration shown in Fig.\,\ref{fig:model}. Based on Knudsen flow approximation, the incident flux of mobile adsorbed species on the surface of the sample at position given by $\mathbf{r}=\rho\hat{x}+Z\hat{z}$ per unit area per unit time is given as

\begin{equation}
\begin{aligned}
&F(s,\phi,\rho,t)\\
&=F_0V(s,\phi,\rho,t)\frac{[\hat{n}_{\mathrm{nozzle}}\cdot(\mathbf{r}-\mathbf{r}')][\hat{n}(\rho,t)\cdot(\mathbf{r}'-\mathbf{r})]}{|\mathbf{r}-\mathbf{r}'|^4}
\end{aligned}
\label{eq:Knudsen}
\end{equation}

Here $F_0$ is a constant that depends on the overall flow rate through the nozzle, and $V(s,\phi,\rho,t)$ is a shadowing parameter computed using a ray-tracing based line of sight estimation from source coordinates $(s,\phi)$ to the surface point at radius $\rho$. Here $V=0$ indicates regions where the flux from the area element does not reach the substrate, and $V=1$ indicates existence of a line of sight.

Simplifying equation (2) we get,

\begin{equation}
\begin{aligned}
&F(s,\phi,\rho,t)\\
&=F_0V(s,\phi,\rho,t)\frac{(h-Z(\rho,t))(s\cos\phi-\rho)n_x(\rho,t)}{(\rho^2+s^2-2s\rho\cos\phi+(h-Z(\rho,t))^2)^2}\\
&+F_0V(s,\phi,\rho,t)\frac{(h-Z(\rho,t))(h-Z(\rho,t))n_z(\rho,t)}{(\rho^2+s^2-2s\rho\cos\phi+(h-Z(\rho,t))^2)^2}
\end{aligned}
\label{eq:Flux_resolved_on_nozzle_elements}
\end{equation}

The total flux at a position $(\rho,Z)$ at time $t$ is therefore expressed by integrating $F$ over the whole circular aperture of the nozzle, i.e.

\begin{equation}
\label{eq:flux_final}
I(\rho,t)=\int_0^{a_{\mathrm{Nozzle}}}s\,ds\int_0^{2\pi}d\phi\,F(s,\phi,\rho,t)
\end{equation}

Note that $I(\rho, t)$ represents the number of molecules incident per unit area per unit time. This term also includes the effect of shadowing (as captured in the line of sight function $V(s,\phi,\rho,t)$ in equation \ref{eq:Flux_resolved_on_nozzle_elements}) and the effect of the surface tilt (as captured via the $[\hat{n}(\rho,t)\cdot(\mathbf{r}'-\mathbf{r})]$ factor in equation \ref{eq:Knudsen}). With this, next we proceed to include the effect of surface diffusion in the growth process.

\subsection{Diffusion model}

We assume the following scenario for our model. An arriving precursor molecule (rate given by equation \ref{eq:flux_final}) is adsorbed on the growth surface, following which it migrates on the surface till it either desorbs from the surface (at a rate $k_{\mathrm{des}}$) or dissociates with the metal incorporating into the growing crystal (rate $k_{\mathrm{inc}}$). For modeling the CVD growth, we assume a simple continuum growth model with purely diffusive mobile adsorbed species migration on the substrate denoted by a single diffusion constant $D$ that is independent of adsorbed species concentration on the surface.  We also ignore any surface curvature driven forces for precursor motion on the surface. The diffusion length is an estimate of the root mean square displacement of the adsorbed precursor molecule over its effective lifetime $t_{\mathrm{tot}}$ = $1/k_{\mathrm{tot}}$, with $k_{\mathrm{tot}} = k_{\mathrm{des}} + k_{\mathrm{inc}}$, representing the net desorption + incorporation rate. If the surface density of the mobile adsorbed species is expressed as $N(\rho,t)$, then the following dynamical rate equation holds:

\begin{equation}
\label{eq:diffusion_equation}
\partial_tN=-k_{\mathrm{tot}}N+D\nabla_S^2N+I(\rho,t)
\end{equation}

Here $\nabla_S^2$ represents the surface Laplacian to account for diffusion along the 2D axisymmetric surface defined by $Z(\rho,t)$ and can be expressed as

\begin{equation}
\label{eq:suface_laplacian}
\begin{aligned}
&\nabla_S^2\cdot\\
&=\left(\frac{1}{\rho\sqrt{1+(\partial Z/\partial\rho)^2}}\right)\partial_\rho\left[\left(\frac{\rho}{\sqrt{1+(\partial Z/\partial\rho)^2}}\right)\partial_\rho\cdot\right]
\end{aligned}
\end{equation}

By invoking the steady-state condition applicable for our regime of growth, we set $\partial_tN=0$, resulting in the equation

\begin{equation}
\label{eq:quasiequilibrium}
\left[1-\frac{D}{k_{\mathrm{tot}}}\nabla_S^2\right]N=\frac{I(\rho,t)}{k_{\mathrm{tot}}}
\end{equation}

The parameter $\frac{D}{k_{\mathrm{tot}}}$ represents the mean square displacement due to the diffusive process at the timescale $\frac{1}{k_{\mathrm{tot}}}$ as captured in equation \ref{eq:diffusion_equation}. We express the diffusion length as 
\begin{equation}\label{num_dem}
    L_{\mathrm{d}}=\sqrt{\frac{D_{s}}{k_{\mathrm{tot}}}}.
\end{equation}.

From equation \ref{eq:quasiequilibrium}, we can express the quasi-equilibrium mobile adsorbed species density on the surface as
\begingroup
\setlength{\abovedisplayskip}{4pt}
\setlength{\belowdisplayskip}{4pt}
\setlength{\abovedisplayshortskip}{2pt}
\setlength{\belowdisplayshortskip}{2pt}
\begin{equation}
\label{eq:quasiequilibrium_inverted}
N(\rho,t)=\frac{1}{k_{\mathrm{tot}}}\left[1-L_{\mathrm{d}}^2\nabla_S^2\right]^{-1}I(\rho,t)
\end{equation}
\endgroup

The normal growth rate is proportional to the mobile adsorbed species density with the proportional constant being the incorporation rate, and can be expressed as

\begin{equation}
\label{eq:normal_growth_rate}
R=k_{\mathrm{inc}}N(\rho,t)=\frac{k_{\mathrm{inc}}}{k_{\mathrm{tot}}}\left[1-L_{\mathrm{d}}^2\nabla_S^2\right]^{-1}I(\rho,t)
\end{equation}

We numerically evaluate $R$ in equation \ref{eq:normal_growth_rate} using a finite difference method and estimate the surface growth in normal direction for small time steps of $\Delta t$. We choose a computational cell with a radius $\rho_{max}>>L_d$ to eliminate any significant boundary effect and set a Dirichlet zero-boundary condition at the computational cell edge. The point on the surface $(\rho,Z(\rho,t))$ moves by a growth vector of $R\Delta t\,\hat{n}(\rho,t)$. This evolution is denoted by updating the vectors corresponding to $\rho$ and $Z$ as
\begin{equation}
\label{eq:timestep}
\begin{aligned}
\rho \ &\mapsto \ \rho +  R\Delta t\ n_x(\rho,t)\\
Z \ &\mapsto \ Z +  R\Delta t\ n_z(\rho,t)\\
\end{aligned}
\end{equation}

After the timestep, the evolved profile is interpolated to a regular $\rho$-grid, and the flux $I(\rho,t)$ is recomputed. The process is repeated till the desired maximum deposition amount, say $Z_{max}$ is reached. We denote the result by $Z^{(Model)}[h,L_{\mathrm{d}},Z_{max}](\rho)$, the computed axis-symmetric profile for specific chosen free parameters: nozzle height $h$, the effective diffusion length parameter $L_{\mathrm{d}}$ that includes diffusion constant $D$ and the overall rate of evaporation + incorporation $(k_{\mathrm{tot}})$, and the maximum deposition amount $Z_{max}$ at $\rho=0$.  
This enables study of the phase space of behavior and comparison with the experimentally measured profiles to estimate the range of the key parameter $L_{\mathrm{d}}$ which gives insight into the diffusion processes.

\subsection{Estimation of $L_{\mathrm{d}}$}\label{eld}

To estimate the effective surface diffusion length, $L_{\mathrm{d}}$, we compare the experimentally obtained deposition profiles with predictions from the model as noted above. The experimental profile is obtained using optical profilometry that provides a three-dimensional height map, $z(x,y)$ over a set of $x$ and $y$ coordinates, $\{x_i,y_i\}$ in a 2D grid of step size $\Delta x = \Delta y = 120\,\mathrm{nm}$. The profile is determined by doing a $z$-scan with a step of $\sim 2\,\mathrm{nm}$ and finding the peak position of the reflected intensity as a function of $z$. The measurement error in $z$, limited by the depth of field, is known to be $\sigma_z\approx 150\,\mathrm{nm}$. More details on the surface profile measurements can be found in Supporting Information section S1. From the measured height-values $z_i = z(x_i, y_i)$, we first extract the center by computing the height-weighted center of mass given by $x_0 = \sum_i x_i z_i /\sum_iz_i$ and $y_0 = \sum_i y_i z_i /\sum_iz_i$. Using this center, the result from the model, $Z[h,L_{\mathrm{d}},Z_{max}](\rho)$ is mapped to the same position grid, i.e. 
\begin{equation}
\label{z_i_model}
z^{(Model)}_i [h,L_{\mathrm{d}}, Z_{max}] = Z[h,L_{\mathrm{d}},Z_{max}]\left(\rho = \rho_i\right)    
\end{equation}
where $\rho_i = \sqrt{(x_i-x_0)^2 + (y_i-y_0)^2}$. We choose the radius of the computation cell $\rho_{max} = 300\mu m$, more than $5L_d$ for the relevant range of diffusion length in our system. This helps eliminate any potential boundary effect affecting the simulated profile. We now compare the measured data $z_i$ with the prediction from the model $z^{(Model)}_i [h,L_{\mathrm{d}}, Z_{max}]$ as a function of the free parameters $h$, $L_{\mathrm{d}}$, and $Z_{max}$. For simplicity we fix $h$ to the experimentally measured nozzle height and find the optimal $L_{\mathrm{d}}$ and $Z_{max}$ by determining the minima in the root mean square error $RMSE[L_{\mathrm{d}}, Z_{max}] = \sqrt{\frac{1}{N}\sum_i |z_i - z^{(Model)}_i [h = h_{exp},L_{\mathrm{d}}, Z_{max}]|^2 }$. To determine the confidence bound for estimation of $L_{\mathrm{d}}$ we calculate its Cramer-Rao bound [\cite{toledo_chapter_2020}] given by

\begin{equation}
    \label{eq:CR_bound_Ld}
    \sigma_{L_{\mathrm{d}}} = \frac
    {\sigma_z}
    {
    \sqrt{\sum_i
    \left\lvert \frac{\partial z_i^{(Model)}}{\partial L_{\mathrm{d}}} \right\rvert ^2}
    }.
\end{equation}
\\

\section*{Acknowledgements}
This work was supported by the U.S. Department of Energy (DOE), Office of Science, Basic Energy Sciences, as part of the the Extreme Lithography and Materials Innovation Center (ELMIC), a Microelectronics Science Research Center (MSRC) under Contract No. DE-AC0206CH11357.  This work made use of the Pritzker Nanofabrication Facility, part of the Pritzker School of Molecular Engineering at the University of Chicago, which receives support from Soft and Hybrid Nanotechnology Experimental (SHyNE) Resource (NSF ECCS-2025633), a node of the National Science Foundation’s National Nanotechnology Coordinated Infrastructure [RRID: SCR\_022955]. We acknowledge the MRSEC Shared User Facilities at the University of Chicago (NSF DMR-2011854). We are grateful for the support of the University of Chicago’s Research Computing Center for assistance with the calculations carried out in this work. Work performed at the Center for Nanoscale Materials, a U.S. Department of Energy Office of Science User Facility, was supported by the U.S. DOE Office of Basic Energy Sciences, under Contract No. DE-AC02-06CH11357. We thank G. Olack from the Department of the Geophysical Sciences at the University of Chicago for the assistance with scanning electron microscopy.

\printbibliography{}

\newpage

\section{Supporting Information Direct-Write Chemical Vapor Deposition}

\subsection{Laser Confocal Microscope measurements}
\begin{figure*}[h]
    \includegraphics[width=0.9\textwidth]{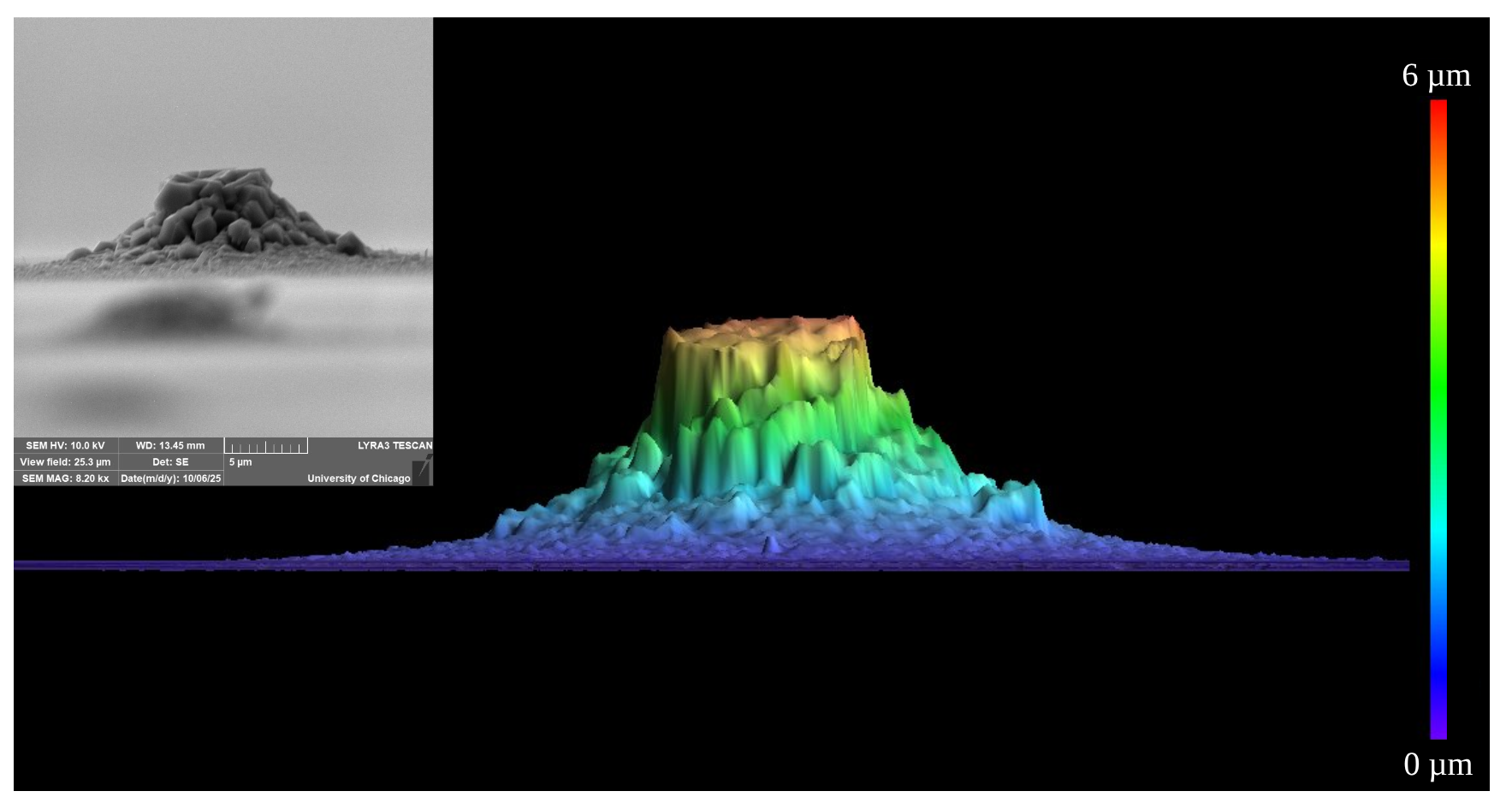}
    \caption{Comparison between direct imaging technique SEM (inset) and 3D plot of the data obtained using laser confocal microscopy. Color bar represents deposit height}
    \label{fig:comp}
\end{figure*}

Following deposition of the matrix of Al spots, the 3D deposit profiles and root mean square surface roughness ($S_{\mathrm{q}}$) are measured using laser confocal microscopy on the Olympus (Evident Scientific) LEXT OLS5000 with a 405 nm laser at 100\textsf{X} magnification, a field of view of 128$\mu$m $\times$ 128$\mu$m. The XY-stage pitch was set to fine mode (0.12 $\mu$m steps) and the Z-stage step height was set to the default value of 5\,$\mathrm{nm}$ to determine a more precise height profile. To confirm the accuracy of our morphological measurements, we deposit a more complex structure by reducing $h$ to 5$\mu\mathrm{m}$ at $185^{\circ}\mathrm{C}$ on a fresh $\mathrm{TiN}$ substrate keeping pressure conditions fixed for ~$t=600~\mathrm{s}$ until an electrical touchdown signal was detected. For nozzle heights smaller than the nozzle diameter, material that has already been deposited as a mound tends to obstruct the trajectory of precursor flux ejected from the nozzle orfice, restricting growth into a distinct top-hat like morphology. Our laser confocal microscope images are able to accurately measure such a complex morphology when compared with a direct imaging technique like SEM, the results of which have been depicted in Fig. \ref{fig:comp}. 

\subsection{SEM images}
\begin{figure}[htbp]
    \centering
    \includegraphics[width=0.9\linewidth]{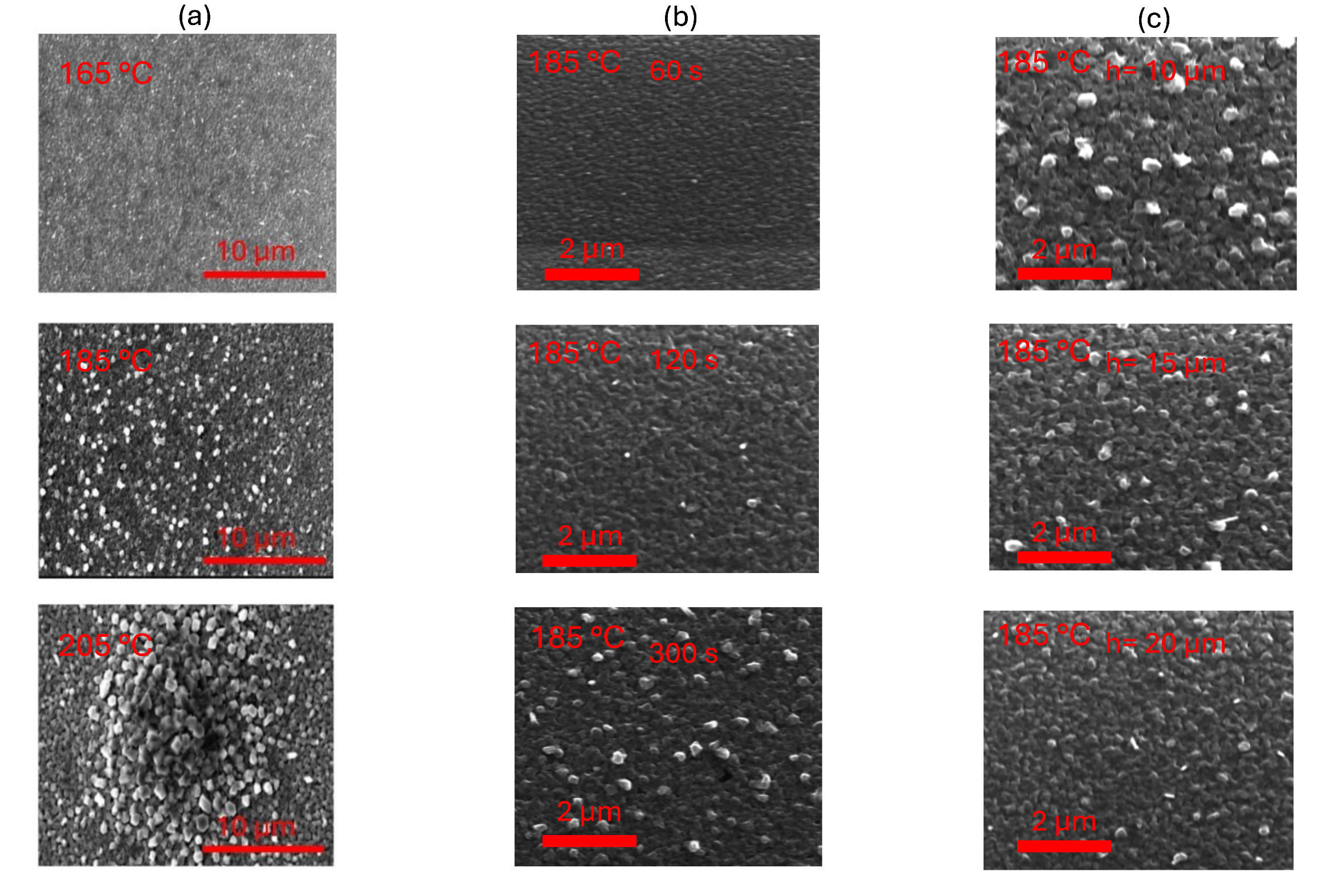}
    \caption{(a) Images of the cs1 deposit microstructure (from consistency scan data) at deposition temperatures of $165^{\circ}C$ (top row), $185^{\circ}C$ (middle row) and $205^{\circ}C$ (bottom row). (b) Images of evolution of the deposit microstructure as a function of deposition time at $185^{\circ}C$ for deposition times of : 60 s (top row), 180 s (middle row) and 300 s (bottom row). All deposits here belong to the time scan column. (c) Images of evolution of the deposit microstructure as a function of nozzle height at $185^{\circ}C$ for nozzle heights of : $h=10\,\mu m$ (top row), $h=15\,\mu m$ (middle row) and $h=20\,\mu m$ (bottom row). All deposits here belong to the height scan column.}
    \label{SEM}
\end{figure}

SEM images of the deposit microstructure as a function of growth temperature, deposition time and nozzle height have been depicted in Fig. \ref{SEM}(a)-(c).

\subsection{Additional Modelling}

\subsubsection{Mean free path estimation}
 We estimate the mean free path ($\lambda$) using the expression $\lambda = \frac{k_{B}T}{\sqrt{2}\pi d_{c}^{2}P}$. The experimental Temperature $T = 363.15\,K$, collision diameter of 0.5 M DMEAA in toluene $d_{c} = 0.573\,\mathrm{nm}$ (calculated using the formalism described in \cite{sharipov2014theoretical,alvarez2013cartography}), and pressure $P = 2e^{-5}\,\mathrm{Torr}$, results in $\lambda = 1.29\,\mathrm{m}$. Since $\lambda \gg h$, where $h$ is our nozzle-substrate separation distance, our system is in the Knudsen flow regime. 

\subsubsection{MOLFLOW simulations}
\begin{figure}[htbp]
    \centering
    \includegraphics[width=0.9\linewidth]{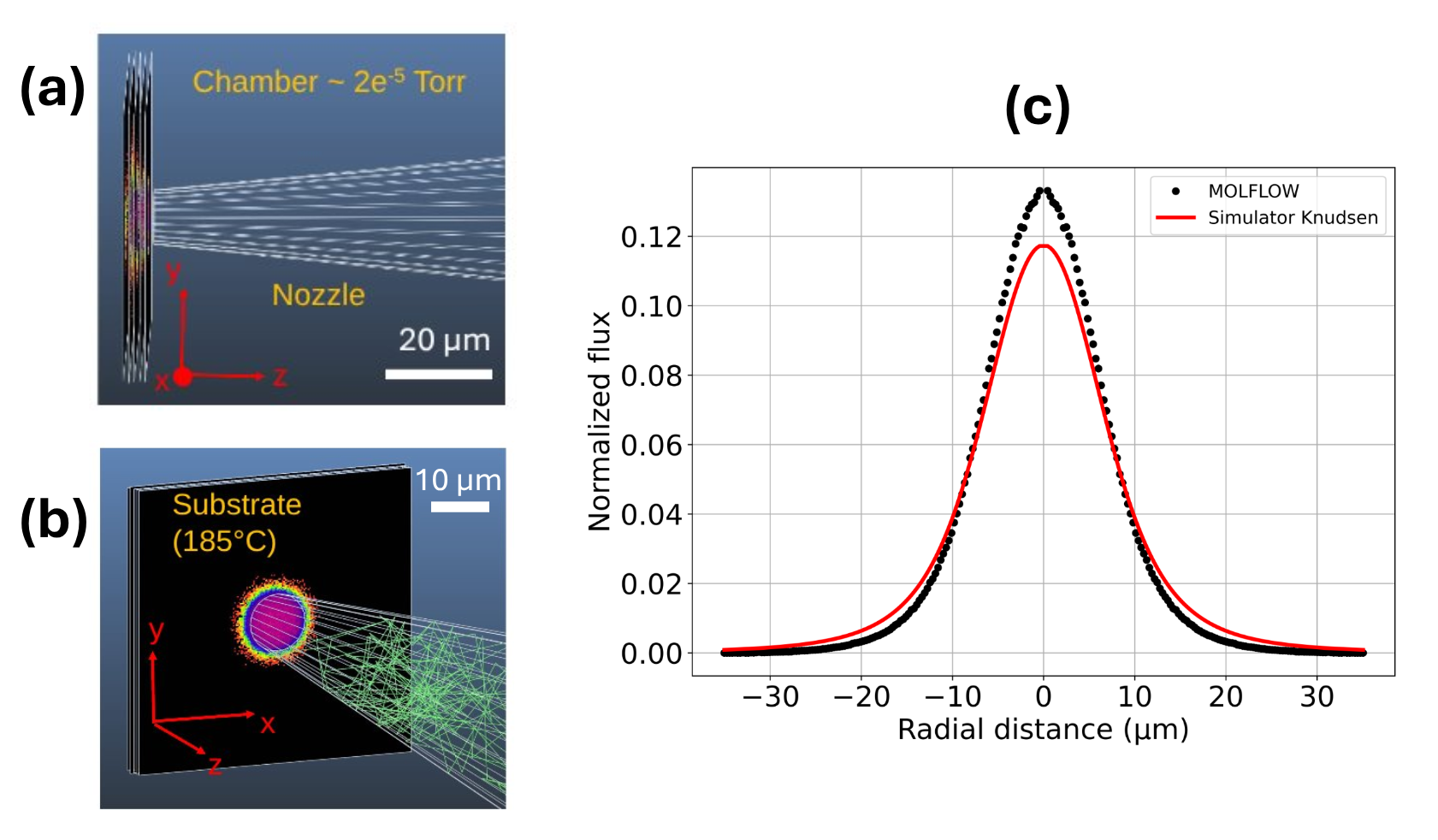}
    \caption{(a) Image of the model geometry including the spatial Precursor flux distribution captured on elemental surfaces within the nozzle-substrate gap region (for $h=10\,\mu m$). (b) Image of the molecular trajectories obtained at the nozzle tip on with the flux incident on an elemental surface right outside the nozzle tip( c) Area Normalized comparison between the precursor flux on the substrate surface ($h = 0~\mu\mathrm{m}$) for the analytical Knudsen flux kernel and MOLFLOW simulation for the simplified nozzle-tip geometry. The comparison serves as a qualitative consistency check.}
    \label{mo}
\end{figure}

To validate the assumption, Molecular-flow simulations were performed in MOLFLOW\cite{kersevan2009introduction,kersevan2019recent} as a qualitative consistency check of the analytical Knudsen flux kernel. The simulation was restricted to the space encompassing the nozzle-tip region and substrate surface (See Fig.\ref{mo}(a)) because the comparison was intended to test molecular spreading after precursor emission into the chamber onto the substrate surface, rather than the full upstream pressure drop through the nozzle. During deposition, the pressure outside the nozzle is approximately $2\times10^{-5}~\mathrm{Torr}$, placing the external transport region in the molecular-flow regime where intermolecular collisions are negligible and transport is governed primarily by the angular distribution of molecules leaving the nozzle and line-of-sight propagation.
The nozzle tip diameter was set to 10 $\mu\mathrm{m}$ to match the experiment. The nozzle temperature was fixed at $90~^\circ\mathrm{C}$, matching the experimental delivery condition. Molecule-wall interactions were treated using non-adsorbing molecular-flow boundary conditions, and the inlet throughput was set to $2.0\times10^{-3}~\mathrm{Torr\,L\,s^{-1}}$. The resulting two-dimensional flux map obtained on the substrate surface was converted into a one-dimensional radial profile by radial averaging about the flux-weighted center of mass. We account for volume conservation across the flux map of the MOLFLOW results for accurate comparison between the Knudsen-simulation and MOLFLOW output. To enforce volume conservation between the theoretical Knudsen profile and the reference MOLFLOW data, a scaling factor is calibrated such that the total integrated axisymmetric volume are equivalent:
\begin{equation}
    V_{\text{Knudsen}} = 2\pi \int_{0}^{r_{\max}} z_{\text{Knudsen}}(r) \, r \, dr = 2\pi \int_{0}^{r_{\max}} z_{\text{MOLFLOW}}(r) \, r \, dr = V_{\text{MOLFLOW}},
\end{equation}

where $z_{\text{Knudsen}}(r)$ and $z_{\text{Molflow}}(r)$ represent the respective radial height distributions, and $r$ denotes the spatial radial coordinate. $ V_{\text{Knudsen}}$ and $ V_{\text{MOLFLOW}}$ represent the axisymmetric volumes of the Knudsen model and MOLFLOW, respectively.

The flux-weighted center was defined as
\[
x_c=\frac{\sum_i x_i \Phi_i}{\sum_i \Phi_i}, 
\qquad
y_c=\frac{\sum_i y_i \Phi_i}{\sum_i \Phi_i},
\]
where $\Phi_i$ is the simulated flux at location $(x_i,y_i)$. The radial coordinate of each point was calculated as
\[
r_i=\sqrt{(x_i-x_c)^2+(y_i-y_c)^2}.
\]
The radial flux profile was obtained by averaging over annular bins,
\[
\Phi(r_j)=
\frac{1}{N_j}\sum_{i\in \mathcal{B}_j}\Phi_i,
\]
\[
\mathcal{B}_j=
\left\{i:\, r_j-\frac{\Delta r}{2}\leq r_i<r_j+\frac{\Delta r}{2}\right\},
\]
where $N_j$ is the number of simulated flux points in the $j^{\mathrm{th}}$ radial bin.
As shown in Fig.\,\ref{mo}, the analytical Knudsen flux kernel captures the dominant spatial features of the MOLFLOW profile: a symmetric centrally peaked distribution, comparable lateral confinement, and monotonic radial decay away from the nozzle axis. The remaining discrepancy is primarily confined to the low-flux tails, where the analytical profile decays more slowly than the simulated profile. This level of agreement is sufficient for the present analysis because the gas-phase kernel is used only as a physically motivated source term for precursor arrival at the substrate. The experimentally extracted quantity is the diffusion length, $L_{\mathrm{d}}$, which is obtained after accounting for surface-mediated broadening. Therefore, the central requirement for the source kernel is that it shows good qualitative agreement with the central localization of the incident flux, followed by its decay towards the low flux tail. We therefore avoid introducing additional parameters solely to force agreement between the analytical and simulated tails. The MOLFLOW comparison supports the use of the analytical Knudsen kernel as a compact, physically motivated description of the dominant gas-phase flux localization, while the residual tail mismatch is treated as part of the model itself.

\end{document}